\documentclass[aps,12pt,showpacs,showkeys]{revtex4}

\usepackage[english]{babel}
\usepackage{amsmath}
\usepackage{amssymb}
\usepackage{amsbsy}
\usepackage{amstext}
\usepackage{graphicx}
\usepackage{pst-node}
\usepackage{verbatim}

\newcommand{\be}{\begin{eqnarray}}
\newcommand{\ee}{\end{eqnarray}}
\newcommand{\bdm}{\begin{displaymath}}
\newcommand{\edm}{\end{displaymath}}
\newcommand{\ds}{\displaystyle}
\newcommand{\ba}{\begin{array}}
\newcommand{\ea}{\end{array}}
\newcommand{\pa}[1]{\left(#1\right)}
\newcommand{\paq}[1]{\left[#1\right]}

\newcommand{\dpa}{\partial}
\newcommand{\p}{{\bf p}}
\newcommand{\K}{{\bf k}}

\newcommand{\X}{{\bf x}}
\newcommand{\R}{{\bf r}}

\begin{document}
\title{Effective field theory calculation of conservative binary dynamics at third post-Newtonian order}

\author{Stefano Foffa$^{\rm 1}$ and Riccardo Sturani$^{\rm 2,3}$}

\affiliation{$(1)$ D\'epartement de Physique Th\'eorique and Center for Astroparticle Physics, Universit\'e de 
             Gen\`eve, CH-1211 Geneva, Switzerland\\
             $(2)$ Dipartimento di Scienze di Base e Fondamenti, 
             Universit\`a di Urbino, I-61029 Urbino, Italy\\
             $(3)$ INFN, Sezione di Firenze, I-50019 Sesto Fiorentino, Italy}

\email{stefano.foffa@unige.ch, riccardo.sturani@uniurb.it}

\begin{abstract}
We reproduce the two-body gravitational conservative dynamics at third 
post-Newtonian order for spin-less sources by using the 
effective field theory methods for the gravitationally bound two-body system,
proposed by Goldberger and Rothstein. 
This result has been obtained by automatizing the computation of Feynman 
amplitudes within a Mathematica algorithm, paving the way for higher-order
computations not yet performed by traditional methods.
\end{abstract}



\maketitle

\section{Introduction}

The problem of finding the equations of motion of a two-body system subject to 
gravitational interaction has been intensively studied since the 
advent of general relativity. Because no exact solution is known, 
different frameworks and approximations have been 
developed out in the past: the present work lies within the post-Newtonian (PN)
approximation of general relativity, see e.g. \cite{Blanchet_living} and \cite{Futamase}
for reviews and current situation of this approach.

In particular the computation of the effective two-body dynamics within the PN 
approximation has 
been the subject of intensive research in the last three decades, both in 
the conservative and in the dissipative sector.
At present the Arnowitt-Deser-Misner Hamiltonian ruling the conservative 
dynamics of a gravitationally bound, spin-less binary system, has been computed
up to third PN order in \cite{energy_at_3PN}, 
whose calculations have been later finalized and confirmed by
\cite{energy_at_3PNHarm,deAndrade:2000gf,Blanchet:2003gy} in harmonic coordinates, and
also by \cite{energy_at_3PN_conf}, where a different resolution of the source
singularity has been adopted.

From the 2.5PN order onwards, the dynamics of binary systems is modified by
dissipative effects \cite{dissip_3.5PN}, associated with gravitational radiation. 
At present the energy flux emitted by a spin-less binary system has been 
computed up to 3.5PN order with respect to the leading order contribution
\cite{flux_3.5PN}.

The main result of this paper is the re-computation of the two-body effective 
action for the conservative dynamics at 3PN order for spin-less objects via
an \emph{automatized algorithm}, making use of
a different method than those implemented so far to obtain such a result:
we performed the computation within the framework of the Effective Field 
Theory methods (EFT) for non relativistic particles introduced in 
\cite{Goldberger:2004jt}. This method has already been applied with success on the conservative sector of the theory
to determine the two-body 1PN \cite{Goldberger:2004jt} and 2PN \cite{Gilmore:2008gq} Hamiltonian (see also \cite{Kol:2009mj} for some investigations beyond 2PN), as well as to the study of the n-body 2PN
\cite{Chu:2008xm} Hamiltonian (for 3-body systems, first obtained in \cite{nbody}).
On the radiative side, \cite{Goldberger:2009qd} showed an application of EFT 
methods to the calculation of some terms of the gravitational wave 
energy flux for spin-less systems up to 3PN order, and in \cite{Cannella:2009he}  it is shown how present gravitational
wave observations can set bounds on the fundamental parameters of the theory.
The EFT approach has lead not only to re-derivation of old and established results, but also
contributed to the computation of new spin effects in the 3PN 
conservative \cite{EFT-E-spin}  (also obtained in \cite{spin-traditional} with more traditional methods) and non-conservative dynamics 
\cite{EFT-flux-spin}.

An EFT method possesses several features making it suitable for computing
physical quantities in problems involving well-separated scales.
This is the case for gravitationally bound two-body systems, where
the different scales can be indentified in the size of the compact
objects $r_s$, the orbital separation $r$ and the gravitational wave-length 
$\lambda$ with hierarchy $r_s < r \sim r_s/v^2 < \lambda \sim r/v$,
being $v$ the relative velocity between the two bodies (we posit
the speed of light $c=1$).
We neglect here radiation effects and consider point-like sources, 
comforted by the effacement principle \cite{effacement}, which 
guarantees that finite size-effects come into play from 5PN order, for 
spin-less objects.
The power-counting scheme outlined in \cite{Goldberger:2004jt}
enables to clearly separate scales and to frame divergences arising from 
zero-size
sources within a well under control field theory setting. Moreover EFT methods
make systematic use of Feynman diagrams and exhibits manifest power counting
rules, enabling to implement PN computations
within an algorithm automatizing the intricate computations
involving hundreds of Feynman diagrams.
Such an algorithm is completely general and it can be 
generalized to the calculation of higher-order dynamics.

A key simplification enabled by EFT methods with respect to the standard ones
is the use of perturbative techniques to integrate out the degrees of freedom 
responsible of mediating the conservative gravitational interaction 
(the so-called 
\emph{potential gravitons} in the language of \cite{Goldberger:2004jt})
to directly obtain the 2-body action, without the need to solve for the metric.

This field of studies is nowadays of great phenomenological impact as the large
interferometers LIGO and Virgo have been operating until recently (fall 2010)
at unprecedented 
sensitivity and a further increase in sensitivity is scheduled for their 
\emph{advanced} runs, due to start in the year 2014, see e.g. 
\cite{:2010yba} for a recent publication of coalescing binary signal 
search. As reported in \cite{:2010cf},
reasonable astrophysical estimates make plausible the detection of 
gravitational wave signals from coalescing binaries by the 
advanced interferometers.

In the search for signals from coalescing binaries the detector output is 
processed via standard match-filtering methods 
\cite{Helmstrom}, where the experimental data are confronted against banks of 
\emph{template} waveforms. The result is particularly sensitive to the time 
varying phase of the oscillating gravitational wave signal 
\cite{Cutler:1992tc}.
In order to compute such a phase with O(1) accuracy it is necessary to take into
account PN corrections to both the energy and the emitted energy flux up to 
3PN order at least.

The frequency region accessible to the advanced versions of LIGO/Virgo (roughly
from $10$Hz to $10$ kHz) corresponds to the last stages (whose duration is 
at most few tens of minutes)
of the coalescence of astrophysical objects like black holes and neutron stars
and by then the orbit will have circularized, as eccentricity decay faster
than orbital separation \cite{Peters:1963ux}, enabling the 
the energy and flux functions to depend on the single parameter $v$.

The inspiral phase of the coalescence is bound to end when the two constituents
of the binary system merge into a single object and the perturbative PN 
analysis cannot be followed beyond the merger. Here however numerical 
techniques have been proven useful, see e.g. \cite{nr_review} for reviews, 
to construct complete analytical waveforms from inspiral 
to merger and the subsequent \emph{ring-down}, as it is conventionally named 
the last phase in which the final object produced by the merger undergoes 
damped oscillation before settling down to quietness. The ring-down
phase also admits an 
analytical perturbative description \cite{ring-down}.

The present work represents a decisive step towards the computation 
of the Hamiltonian dynamics at fourth PN order, which is a key 
ingredient in the construction of template waveforms to be used in actual
experimental searches, like the the Effective One Body waveforms
\cite{EOB} which have been used in the recent \cite{Abadie:2011kd}.

The paper is organized as follows. In sec.~\ref{main} we give an overview
of the effective
field theory methods for gravity, which are applied in sec.~\ref{results} to the
determination of the 2-body effective action at order 3PN. We summarize and 
conclude in sec.~\ref{conclusion}.

\section{Main}
\subsection{Lagrangian and Feynman rules}
\label{main}
The relevant physical scales for the 2-body problem in gravity are the total 
mass $M$, the separation $r$ and the relative velocity $v$.
By nPN correction it is conventionally meant corrections of order
$v^{2n}\sim (G_N M/r)^n$, where $G_N$ is the standard Newton constant and 
the virial relation has been applied,
showing that an expansion in $v^2$ is at the same time an expansion in the
strength of the gravitational field.

At 3PN order, the EFT description of massive compact objects in binary systems takes them as 
non-dynamical, background point-like sources: quantitatively this corresponds 
to particle worldlines interacting with gravitons. 
The action $S$ we consider is then given by 
\be
\label{action}
S = S_{EH}+S_{GF}+S_{pp}\,,
\ee
where the first and the third terms are, respectively,
the $d$-dimensional Einstein-Hilbert action
and the worldline point particle action
\be
\label{az_EH}
\ba{rcl}
\ds S_{EH} &=&\ds 2 \Lambda^2\int {\rm d}^{d+1}x\sqrt{-g}\ R(g)\,,\\
\ds S_{pp} &=&\ds -\sum_{i=1,2} m_i\int {\rm d}\tau_i = 
-\sum_{i=1,2} m_i\int \sqrt{-g_{\mu\nu}(x^\mu_i) {\rm d}x_i^\mu {\rm d}x_i^\nu}\,,
\ea
\ee
with 
$\Lambda^{-2}\equiv 32\pi G$, being $G$ the $d$-dimensional gravitational constant\footnote{We adopt the ``mostly plus'' convention:
$\eta_{\mu\nu}\equiv {\rm diag}(-,+,+,+)$; the Riemann and Ricci tensors are
defined as $R^\mu_{\nu\rho\sigma}\equiv\dpa_\rho\Gamma^\mu_{\nu\sigma}+
\Gamma^\mu_{\alpha\rho}\Gamma^\alpha_{\nu\sigma}-\rho\leftrightarrow\sigma$, with $\Gamma^\mu_{\alpha\beta}$ the usual Christoffel symbol and $R_{\mu\nu}\equiv R^\alpha_{\mu\alpha\nu}$. }:
as dimensional regularization will be 
needed, it is necessary to keep $d$ generic until the actual value of
Feynman integrals will have been computed.
The goal is to compute the effective action $S_{eff}$ for particles alone, with 
gravitons mediating interactions being integrated out by standard perturbative methods, i.e.
by the aid of Feynman diagrams. 
For the gauge fixing term, we follow \cite{Gilmore:2008gq}:
\be
S_{GF}=- \Lambda^2\int {\rm d}^{d+1}x \sqrt{-g}\Gamma_\mu\Gamma^\mu
\ee
with $\Gamma^\mu\equiv \Gamma^\mu_{\alpha\beta}g^{\alpha\beta}$,
corresponding to the same harmonic gauge adopted in \cite{Blanchet_living}.
Still following \cite{Gilmore:2008gq}, we adopt the standard Kaluza-Klein (KK) 
parametrization  of the metric \cite{Kol:2010si} (a somehow similar 
parametrization was first applied within the framework of a PN calculation in
 \cite{annales}), suitable for a nonrelativistic expansion around a 
Minkowski metric:
\be
\label{met_nr}
g_{\mu\nu}=e^{2\phi/\Lambda}\pa{
\ba{cc}
-1 & A_j/\Lambda \\
A_i/\Lambda &\quad e^{- c_d\phi/\Lambda}\gamma_{ij}-
A_iA_j/\Lambda^2\\
\ea
}\,,
\ee
with $\gamma_{ij}=\delta_{ij}+\sigma_{ij}/\Lambda$,
$c_d=2\frac{(d-1)}{(d-2)}$ and $i,j$ running over the $d$ spatial dimensions.

In terms of the metric parametrization (\ref{met_nr}),
each world-line coupling to the gravitational degrees of freedom
$\phi$, $A_i$, $\sigma_{ij}$  reads
\renewcommand{\arraystretch}{1.4}
\be
\label{matter_grav}
S_{pp}=-m\ds \int {\rm d}\tau = \ds-m\int {\rm d}t\ e^{\phi/\Lambda}
\sqrt{\pa{1-\frac{A_i}{\Lambda}v^i}^2
-e^{-c_d \phi/\Lambda}\pa{v^2+\frac{\sigma_{ij}}{\Lambda}v^iv^j}}\,,
\ee
\renewcommand{\arraystretch}{1.}
and its Taylor expansion provides the various particle-gravity vertices of the 
EFT.

Also the pure gravity sector $S_{bulk}=S_{EH}+ S_{GF}$ can be explicitly written
in terms of the KK variables, $S_{EH}$ in KK variable has been first derived in
\cite{Kol:2010si}. We report here only a part of $S_{bulk}$, including all the 
terms that are needed for the 3PN calculation:
\renewcommand{\arraystretch}{1.4}
\be
\label{bulk_action}
S_{bulk}&\supset&\int {\rm d}^{d+1}x\sqrt{-\gamma}
\left\{\frac{1}{4}\left[(\nabla\sigma)^2-2(\nabla\sigma_{ij})^2-\left(\dot{\sigma}^2-2(\dot{\sigma}_{ij})^2\right){\rm e}^{\frac{-c_d \phi}{\Lambda}}\right]
- c_d \left[(\nabla\phi)^2-\dot{\phi}^2 {\rm e}^{-\frac{c_d\phi}{\Lambda}}\right]\right.\nonumber\\
&&
+\left[\frac{F_{ij}^2}{2}+\left({\bf{\nabla}}\!\!\cdot\!\!{\bf{A}}\right)^2 -\dot{A}^2 {\rm e}^{-\frac{c_d\phi}{\Lambda}} \right]
{\rm e}^{\frac{c_d \phi}{\Lambda}}+2 c_d \left(\dot{\phi}{\bf{\nabla}}\!\!\cdot\!\!{\bf{A}}-\dot{{\bf{A}}}\!\!\cdot\!\!{\bf{\nabla}}\phi\right)\nonumber\\
&&
+\left.\frac{1}{\Lambda}\left(2\left[F_{ij}A^i\dot{A^j}+{\bf{A}}\!\!\cdot\!\!\dot{{\bf{A}}}({\bf{\nabla}}\!\!\cdot\!\!{\bf{A}})\right]{\rm e}^{\frac{c_d\phi}{\Lambda}}-2 c_d\dot{\phi} {\bf{A}}\!\!\cdot\!\!{\bf{\nabla}}\phi\right)\right\}\,,
\ee
\renewcommand{\arraystretch}{1.}
where $F_{ij}\equiv A_{j,i}- A_{i,j}$ and 
indices are raised and
contracted by means the $d$-dimensional metric tensor $\gamma$
(although in most terms one can just use $\delta_{ij}$ because the neglected 
parts are not needed at 3PN)\footnote{We denote $d$-vectors with bold 
characters and their modulus by the standard character,
i.e. for a vector $\bf{B}$ we have 
$B=\pa{\sum_{i=1}^dB_iB_j\gamma^{ij}}^{1/2}=\pa{{\mathbf B}^2}^{1/2}$.}. 
All spatial derivatives
are meant to be simple (not covariant)
\footnote{One should not be surprised by the lack of manifest covariance,
even with respect to the $d$-dimensional metric $\gamma$, as the
introduction of the gauge fixing term breaks general covariance.}
and, when ambiguities might raise, gradients are always meant to act on
contravariant fields (so that, for instance,
${\mathbf{\nabla}}\!\!\cdot\!\!{\mathbf{A}}\equiv\gamma^{ij}A_{i,j}$
and $F_{ij}^2\equiv\gamma^{ik}\gamma^{jl}F_{ij}F_{kl}$).

All the Feynman rules needed for our calculation can be read off eq.~(\ref{bulk_action}).
In particular: the first line gives the $\sigma$ and $\phi$ propagators, as well as all the
$\phi^n \sigma^m$ interaction vertices;
the second line gives the ${\bf A}$ propagator and,
together with the third, contains enough information to compute all the other vertices needed for the 3PN calculation.

Borrowing the definitions from \cite{Goldberger:2004jt}, the two
bodies exchange \emph{potential} gravitons, responsible for binding the system 
as they mediate instantaneous interactions: their characteristic four-momentum 
$k_\mu$ scales thus as $k_\mu\sim (k^0\sim v/r , \K\sim r^{-1})$
and we neglect altogether the emission of gravitational waves because we
are interested in the conservative part of the binary system dynamics.
When a compact object emits a single graviton, momentum is actually 
\emph{not} conserved and the non-relativistic particle recoils of a fractional 
amount roughly given by $|\delta{p}|/|p|\simeq |k|/|p|\simeq \hbar/J$, 
where $J\sim mvr$ is the angular momentum of the system: 
however, for macroscopic systems such a quantity is negligibly small.

In order to allow manifest scaling it is necessary to work with the
space-Fourier transformed fields
\renewcommand{\arraystretch}{1.4}
\be
\label{Fourk}
\ba{rcl}
\phi_\K(t) &\equiv& \ds\int {\rm d}^dx\, \phi(t,\X) e^{-i\K \X}\,,\\
A_{i, \K}(t) &\equiv& \ds\int {\rm d}^dx\, A_i(t,\X) e^{-i\K \X}\,,\\
\sigma_{ij, \K}(t) &\equiv& \ds\int {\rm d}^dx\,\sigma_{ij}(t,\X) e^{-i\K \X}
\,.
\ea
\ee
\renewcommand{\arraystretch}{1}
The fields defined above are the fundamental variables in terms of
which the Feynman graphs are going to be constructed; the action governing
their dynamics can be found from eqns.~(\ref{matter_grav},\ref{bulk_action})
by direct substitution and replacement of spatial derivatives by the appropriate $i\K$ factors.

From the quadratic part of the Lagrangian the propagators can be 
written as:
\renewcommand{\arraystretch}{1.4}
\be
\label{propagators}
\left.
\ba{rcl}
\ds P[\phi_\K(t_a)\phi_{\K'}(t_b)]&=&\ds -\frac{1}{2 c_d}\\
\ds P[A_{i, \K}(t_a)A_{j, \K'}(t_b)]&=&\ds\frac 12 \delta_{ij}\\
\ds P[\sigma_{ij, \K}(t_a)\sigma_{kl, \K'}(t_b)]&=&\ds \frac12 P_{ij,kl}
\ea
\right\}\times (2\pi)^d\delta^{d}(\K+\K'){\cal P}(\K^2,t_a,t_b)\delta(t_a-t_b)\,,
\ee
\renewcommand{\arraystretch}{1}
where $P_{ij,kl}\equiv -\left[\delta_{ik}\delta_{jl}+\delta_{il}\delta_{jk}+(2-c_d)\delta_{ij}\delta_{kl}\right]$\, and
\be
\label{prop_serie}
{\cal P}(\K^2,t_a,t_b)=\frac{i}{\K^2-\partial_{t_a}\partial_{t_b}}\simeq
\frac{i}{\K^2}\left(1+\frac{\partial_{t_a}\partial_{t_b}}{\K^2}+
\frac{\partial_{t_a}^2\partial_{t_b}^2}{\K^4}+\dots\right)
\ee
is the full relativistic propagator,
which can be thought as an instantaneous nonrelativistic
part plus insertion terms involving time derivatives.
Only few of these time derivative terms need to be included in a Feynman
diagram at a given PN order.

\subsection{Feynman diagrams}

We have exposed all the ingredients needed to obtain the 2-body 
effective action
$S_{eff}$ with manifest power counting in $G$ and $v$ at the third
post-Newtonian order.
This can be done by integrating out the graviton fields from the full
action derived above
\begin{equation}\label{pathint}
  S_{eff}=\int D\phi D\sigma_{ij} DA_k\exp[i(S_{EH}+S_{GF}+ S_{pp})]\,.
\end{equation}
We can now proceed to lay down all the Feynman diagrams relevant for our
computation, then by applying the Feynman rules derived from the above
interactions the amplitudes will be computed, collecting all results
belonging to the same PN order.

\subsubsection{Topologies}

Following the procedure and the terminology of \cite{Gilmore:2008gq},
we first select the topologies having the appropriate power of $G$ for being 
potentially relevant from a given PN order on, that is $G^{(n+1)}$ for 
$n$th-PN order.
The attribution of a power of $G$ to a topology is made according to the
following rule: each vertex involving $n$ gravitational fields gives a
factor $G^{n/2-1}$ if it is a bulk one, and a factor $G^{n/2}$ if it is
attached to an external particle.

Concretely, there is just one topology contributing at the Newtonian order
$G$: the one representing the single graviton exchange.
At ${\cal O}(G^2)$ order two other topologies (modulo switch of the
external particles) enter the game
but actually just one of these contributes at 1PN since the other involves an extra
$v^2$ factor, which is equivalent to increasing by one unity the PN order.
As underlined in \cite{Gilmore:2008gq}, this is a very welcome effect of using 
the KK parametrization, also with respect to the Arnowitt-Deser-Misner
one, see \cite{Kol:2010ze}.

\begin{figure}
\includegraphics[width=1.\linewidth]{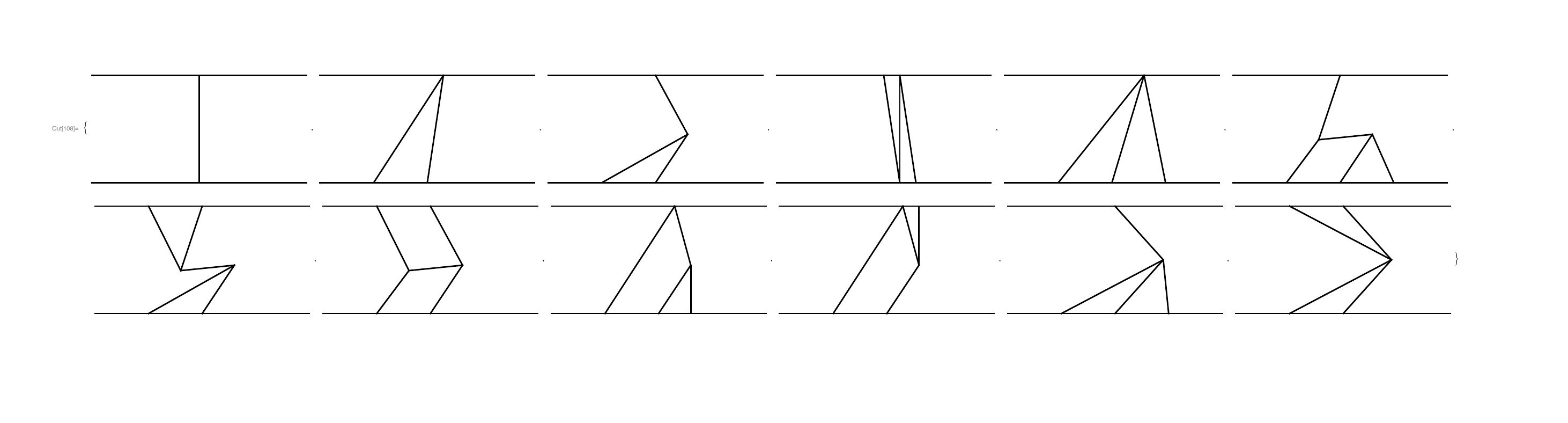}
\caption{The 12 topologies up to $G^3$ as they are produced by a Mathematica
code. The last 4 start contributing at 3PN order. Here and in the following figures, we count them form left to right, from top to bottom.}
\label{topg123}
\end{figure}

At 2PN order, 5 new $G^3$ topologies have to be added, while 4 additional
ones come into play only from 3PN on because of $v^2$ suppressions.
All these 12 topologies contributing to $G^3$ order have been already displayed in \cite{Gilmore:2008gq},
and they are also reported here in figure \ref{topg123}.

To complete the 3PN order the introduction of $G^4$
topologies is required. The generation of the new topologies at any given order
$n$ is done iteratively from the ones at the previous order $n-1$:
to any $G^{n-1}$ topology a new propagator is attached in all possible ways,
provided that one extremum ends on an external particle, and the second
extremum is placed in one of the following three locations: i) an internal vertex, 
ii) a vertex attached to the other external particle, iii) a propagator,
so to "break" it into two propagators ending into the same newly-born 3-legged 
internal vertex; barring  in all cases the possibility that internal graviton 
loops are created, as they would give rise to uninteresting quantum corrections
(i.e. proportional to $\hbar/J\ll 1$).\\ 
In this way one actually obtains a redundant set of
$G^{n}$ topologies because many of them are equivalent, i.e.
they are skeletons of the same Feynman graphs, then the algorithm
to eliminate equivalent graphs can be straightforwardly 
implemented on a computer program.

\begin{figure}
\includegraphics[width=1.\linewidth]{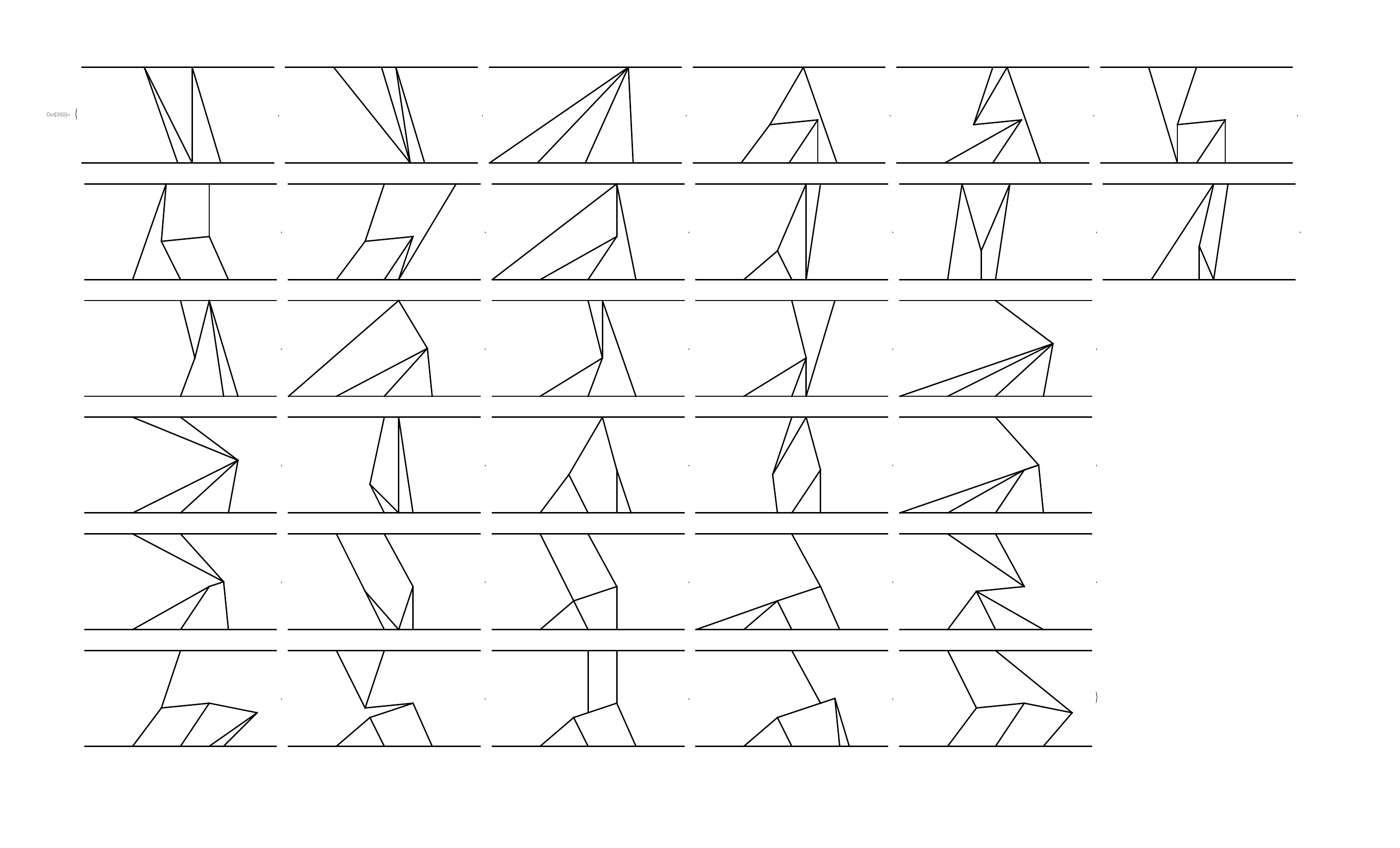}
\caption{The 32 $G^4$ topologies. Only the first 8 contribute at 3PN order, 
and they are all factorisable into simpler subtopologies.}\label{topg4}
\end{figure}

We have indeed implemented the whole procedure into a Mathematica code 
\cite{Mathematica} and 32 $G^4$ topologies have been found, only 8 of which
turned out to be relevant at the 3PN level of accuracy, as the remaining 24
contribute at higher PN-orders because of $v^2$ suppressions.
Moreover, as figure \ref{topg4} shows,
none of the 8 3PN-relevant topologies is really new, since they are actually 
factorisable in lower order sub-topologies.
We have also worked out, for the sake of future developments, the situation at 
4PN and summarized the number of topologies in the left part of table 
\ref{tabdiatop}.

That makes a total of 20 topologies that must be taken into account for the 3PN
calculation, and only 2 of them (the two $G^3$ ones involving 4-legged 
vertices, the last two ones of fig.~\ref{topg123}) 
are genuinely new ones, i.e. non-factorisable into subtopologies.

\begin{table}
\begin{tabular}{l | c  c  c  c  c}
 & 0PN & 1PN & 2PN & 3PN & 4PN\\
\hline
$G$  & 1 &  &  &  &\\
$G^2$ &  & 1 & 1 & &\\
$G^3$ &  &  & 5 & 4 &\\
$G^4$ &  &  &  & 8 & 21\\ 
$G^5$ &  &  &  &  & 50\\
\end{tabular}
\quad\quad\quad\quad\quad\quad
\begin{tabular}{l | c  c  c  c}
 & $v^0$ & $v^2$ & $v^4$ & $v^6$\\
\hline
$G$  & 1 & 1 & 1 & \\
$G^2$ & 1 & 8 & {\bf 7} & 7\\
$G^3$ & 5 & {\bf 48} & 159 & \dots\\
$G^4$ & {\bf 8} & 299 &  \dots &\\
$G^4$ & 50 & \dots &  &\\
\end{tabular}
\caption{On the left: number of topologies entering at a given PN order. 
On the right: number of diagrams that start to contribute to the effective 
action at a given $G$ and $v$ power. The 63 diagrams whose contribution starts 
at 3PN are written in boldface characters.}\label{tabdiatop}
\end{table}

\subsubsection{Graphs}

The next step is to generate Feynman graphs from all topologies  by
replacing all the not-yet-specified propagators by the proper
$\phi$, $A$ or $\sigma$ ones.
This is the point where powers of $v$ (traced by
time derivatives in the Lagrangian) enter the game,
both in external and bulk vertices and in the relativistic propagators.
The lowest power of $v^2$ appearing in a graph, together with its $G$-order,
determines the lowest PN order at which such a graph must be taken into account.
As different vertices carry different time derivative
structures, see for instance eq.~(\ref{bulk_action}),
topologies of the same $G$-order, or even the same topology, can generate
Feynman graphs whose lowest PN orders differ.
In particular some topologies allow only graphs containing
$v^2$ corrections at least, thus entering the game at higher PN order than one 
would have inferred
by a naive $G$-power counting; this is the origin of the above mentioned 
simplification introduced by the use of KK variables,
which reduces the number of topologies (and thus of graphs) to be taken into 
account at any given PN order.
This welcome feature becomes more and more important at higher PN orders.

The generation and classification of all the Feynman graphs needed at a
given PN level is an automatizable procedure,
as it suffices to try all possibilities (although one can envisage
shortcuts towards more efficient algorithms) and find out the
lowest power of $v^2$ of each graph using the Feynman rules
implicit in the Lagrangians (\ref{matter_grav},\ref{bulk_action}).
Our Mathematica code has thus generated all 80 the diagrams giving a 
contribution
at 3PN (as well as other 515 for the 4PN case): 63 of them are new ones, while 
the other 17 have been computed at 2PN order by Gilmore and 
Ross \cite{Gilmore:2008gq}\footnote{Their
accounting gives actually 21 diagrams, but this is due to the fact that we 
consider here two diagrams differing only by the addition of 
a higher-order term in the propagator eq.~(\ref{prop_serie}) as just the 
series expansion of the same diagram. 
Had we used their convention, the number 
of new 3PN diagrams would have been more than one hundred, while the 4PN 
one would be around one thousand.}.
The right part of table \ref{tabdiatop} shows the classification of the new 
diagrams according to their $G$ and lowest $v^2$ powers.

\section{The results}
\label{results}
We use a Mathematica \cite{Mathematica} code together with EinS \cite{EinS} and Feyncalc 
\cite{Mertig:1990an} softwares to generate and to compute all the diagrams.
Here we show the results concerning the 3PN part only
\footnote{After the submission of this work, Michele Levi from the Hebrew University has informed us that she managed to compute,
confirming our results, a subset of the diagrams presented in this section (more precisely, all the diagrams except for $G^3_{5, 34-37, 42,43}$)}.

\subsection{${\cal O}(G)$ diagrams}
They are shown in figure \ref{diaG1} and they can all be done by means of the dimensionally regularized Fourier formula
\be
\label{Fourp}
\ds\int_ {\p}\, \frac{e^{-i\p \R}}{p^{2\alpha}}=\frac 1{\pa{4\pi}^{d/2}}
\frac{\Gamma(d/2-\alpha)}{\Gamma(\alpha)}\pa{\frac 2r}^{d-2\alpha}
\ee
(with $\int_ {\p}\equiv\int \frac{{\rm d}^d p}{(2 \pi)d}$ and $\R\equiv{\bf x_1}-{\bf x_2}$), as well as its generalization to the case where factors of
$p^\mu$'s appear in the integrand; the corresponding formulae can be obtained
by taking the appropriate number of $\R$-gradients of equation (\ref{Fourp}).

The result is then multiplied by a symmetry factor accounting for:
\begin{itemize}
\item factorials coming from the development of the exponential function 
contained in (\ref{pathint});
\item a factor 1/2 for diagrams symmetric under particles exchange, 
in order not to overcount when performing symmetrization (i.e. 
$+\ 1\leftrightarrow2$) in the very final stage of the calculation;
\item the number of possible Wick contractions giving rise to the same diagram,
with the convention that equivalent legs attached to an \emph{internal} vertex 
are indistinguishable
(because we chose to incorporate the corresponding factors already in the 
internal vertices definitions).
\end{itemize}
The results of the integration have been evaluated at $d=3$ and multiplied
 by the imaginary unit $i$ to give the following potential terms for the 3PN effective Lagrangian
($G^{a}_b$ being the value of the $b$-th $G^{a}$ diagram, listed as in figures \ref{diaG1}-\ref{diaG4}):
\begin{center}
  \begin{figure}
    \includegraphics[width=1\linewidth]{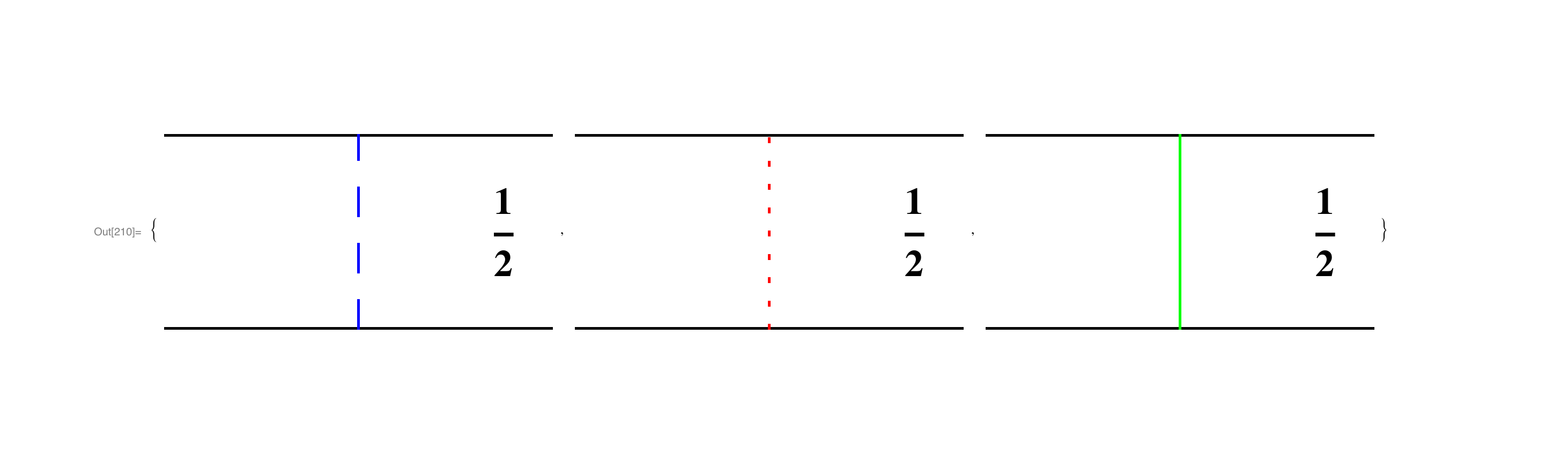}
    \caption{The three ${\cal O}(G)$ diagrams. The $\phi$, $A$ and $\sigma$ 
      propagators are represented respectively by blue dashed, red dotted and 
      green solid lines. From the left to the right, they 
      contain $v^6$, $v^4$ and $v^2$ corrections under the form of vertices 
      expansion and/or double time derivative insertions in the propagators. 
      The number in the right part of each diagram is its multiplicity factor.}
    \label{diaG1}
  \end{figure}
\end{center}
\be
G^{1}_1&=&-\frac{G_N m_1 m_2}{32 r} \left\{\frac{{\bf \dot{a}_1\cdot\dot{a}_2}+3{\bf n.\dot{a}_1} {\bf n.\dot{a}_2}}{9}r^4+\left[{\bf a_1.a_2} {\bf r.a_2} -2 {\bf a_1.a_2} {\bf r.a_1}  - {\bf a_2.\dot{a}_1} {\bf r.v_1} - {\bf a_2.\dot{a}_1} {\bf r.v_2}\right.\right.\nonumber\\
&&-2 {\bf r.\dot{a}_1} {\bf v_1.a_2} -18 {\bf v_1.a_1} {\bf v_1.a_2} -2 {\bf a_1.a_2} v_1^2 +9 a_2^2 v_1^2 -2 {\bf a_1.a_2} {\bf v_1.v_2} - {\bf v_1.a_2} {\bf v_2.a_1} -14 {\bf r.\dot{a}_1} {\bf v_2.a_2}\nonumber\\
&&\left.-111 {\bf v_1.a_1} {\bf v_2.a_2} -9 a_1^2 v_2^2-7 {\bf a_1.a_2} v_2^2\right]r^2
+22 v_1^6+2 ({\bf v_1.v_2})^3-2 {\bf a_1.a_2} ({\bf r.v_1})^2 -{\bf a_1.a_2} ({\bf r.v_2})^2\nonumber\\
&&+16 v_1^2 ({\bf v_1.v_2})^2 - ({\bf r.a_1})^2 {\bf r.a_2}- {\bf r.a_2} {\bf r.\dot{a}_1} {\bf r.v_1} - {\bf r.a_2} {\bf r.\dot{a}_1} {\bf r.v_2} -2 {\bf a_1.a_2} {\bf r.v_1} {\bf r.v_2}\nonumber\\
&&-7 {\bf r.a_1} {\bf r.v_1} {\bf v_1.a_2}-3{\bf r.a_1} {\bf r.v_2} {\bf v_1.a_2} -3 ({\bf r.a_2})^2 v_1^2 -2 {\bf r.a_1} {\bf r.a_2} v_1^2 +88 {\bf r.v_2} {\bf v_1.a_1} v_1^2\nonumber\\
&& +22 v_1^4 {\bf v_1.v_2} -2 {\bf r.a_1} {\bf r.a_2} {\bf v_1.v_2} +32 {\bf r.v_2} {\bf v_1.a_1} {\bf v_1.v_2} +16 {\bf r.v_2} v_1^2 {\bf v_2.a_1}  +8 {\bf r.v_2} {\bf v_1.v_2} {\bf v_2.a_1}\nonumber\\
&&-41 {\bf r.a_1} {\bf r.v_1} {\bf v_2.a_2}-17 {\bf r.a_1} {\bf r.v_2} {\bf v_2.a_2}+3 ({\bf r.a_1})^2 v_2^2 +34 v_1^4 v_2^2 -7 {\bf r.a_1} {\bf r.a_2} v_2^2+124 {\bf r.v_2} {\bf v_1.a_1} v_2^2\nonumber
\ee
\be
&&+23 v_1^2 {\bf v_1.v_2} v_2^2 +20 {\bf r.v_2} {\bf v_2.a_1} v_2^2+2 {\bf r.a_1} {\bf r.a_2} ({\bf n.v_1})^2+{\bf r.a_1} {\bf r.a_2} ({\bf n.v_2})^2 +8 ({\bf n.v_2})^2 v_1^4\nonumber\\
&&  -22 {\bf n.v_1} {\bf n.v_2} v_1^4-6 {\bf n.v_1} {\bf n.v_2} ({\bf v_1.v_2})^2  +2 {\bf r.a_1} {\bf r.a_2} {\bf n.v_1} {\bf n.v_2} -4 {\bf r.a_1} ({\bf n.v_2})^2 {\bf v_1.v_2} \nonumber\\
&&+12 ({\bf n.v_2})^2 v_1^2 {\bf v_1.v_2} -16 {\bf n.v_1} {\bf n.v_2} v_1^2 {\bf v_1.v_2} -8 ({\bf n.v_2})^3 {\bf v_2.a_1} r/3 -4 {\bf r.v_1} ({\bf n.v_2})^2 {\bf v_2.a_1} \nonumber\\
&&-20 {\bf r.a_1} ({\bf n.v_1})^2 v_2^2 -10 {\bf r.a_1} ({\bf n.v_2})^2 v_2^2  -16 {\bf r.a_1} {\bf n.v_1} {\bf n.v_2} v_2^2-20 ({\bf n.v_1})^2 v_1^2 v_2^2\nonumber\\
&&  -39 {\bf n.v_1} {\bf n.v_2} v_1^2 v_2^2+2 {\bf r.a_1} ({\bf n.v_2})^4  +4 {\bf r.a_1} {\bf n.v_1} ({\bf n.v_2})^3 +20 ({\bf n.v_2})^4 v_1^2+4 {\bf n.v_1} ({\bf n.v_2})^3 v_1^2 \nonumber\\
&&\left.+9 ({\bf n.v_1})^2 ({\bf n.v_2})^2 {\bf v_1.v_2}  -5 ({\bf n.v_1})^3 ({\bf n.v_2})^3\right\}\,,\nonumber\\
G^{1}_2&=&\frac{G_N m_1 m_2 }{4 r}\left\{r^4 \frac{\bf \dot{a}_1.\dot{a}_2}{3}+\left[{\bf a_1.a_2} {\bf r.a_2}-2 {\bf a_2.\dot{a}_1} {\bf r.v_1}-2 {\bf v_1.a_2} {\bf r.\dot{a}_1}-2 {\bf a_2.\dot{a}_1} {\bf r.v_2}\right.\right.\nonumber\\
&&\left.-3{\bf a_1.a_2} {\bf v_1.v_2}-2  {\bf v_2.a_1} {\bf v_1.a_2}-14 {\bf v_1.a_1} {\bf v_1.a_2}-3  v_1^2 {\bf a_1.a_2}-5 v_2^2 {\bf a_1.a_2}-3 {\bf a_1.a_2} {\bf r.a_1} \right] r^2\nonumber\\
&&-2 {\bf a_1.a_2} {\bf r.v_1} {\bf r.v_2}-2 {\bf r.a_1} {\bf v_1.a_2} {\bf r.v_2}- {\bf r.a_1} {\bf r.a_2} {\bf v_1.v_2}-3 {\bf a_1.a_2} ({\bf r.v_1})^2-6 {\bf r.a_1} {\bf v_1.a_2} {\bf r.v_1}\nonumber\\
&&- {\bf a_1.a_2} ({\bf r.v_2})^2+12 {\bf v_1.a_1} {\bf r.v_2} {\bf v_1.v_2}+6 v_1^2 {\bf v_2.a_1} {\bf r.v_2}+8 {\bf v_2.a_1} {\bf r.v_2} {\bf v_1.v_2}+10 v_2^2 {\bf v_2.a_1} {\bf r.v_2}\nonumber\\
&&-2 {\bf r.a_1} ({\bf n.v_2})^2 {\bf v_1.v_2}-2 {\bf v_2.a_1} {\bf r.v_1} ({\bf n.v_2})^2-2 r {\bf v_2.a_1} ({\bf n.v_2})^3+2 ({\bf v_1.v_2})^3+6 v_1^2 ({\bf v_1.v_2})^2\nonumber\\
 &&\left.+6 v_1^4 {\bf v_1.v_2}+v_1^2 v_2^2 {\bf v_1.v_2}-4 {\bf n.v_1} {\bf n.v_2} ({\bf v_1.v_2})^2-6 v_1^2 {\bf n.v_1} {\bf n.v_2} {\bf v_1.v_2}+3 ({\bf n.v_1})^2 ({\bf n.v_2})^2 {\bf v_1.v_2}\right\}\,,\nonumber\\
G^{1}_3&=&\frac{G_N m_1 m_2 }{2 r}\left\{\left[2  {\bf a_1.a_2} {\bf v_1.v_2}+2  {\bf v_2.a_1} {\bf v_1.a_2}-4  {\bf v_1.a_1} {\bf v_2.a_2}\right]r^2
\right.\nonumber\\
&&-4 {\bf v_2.a_1} {\bf r.v_2} {\bf v_1.v_2}+4 v_2^2 {\bf v_1.a_1} {\bf r.v_2}-({\bf v_1.v_2})^3-2 v_1^2 ({\bf v_1.v_2})^2+2 v_1^4 v_2^2+v_1^2 v_2^2 {\bf v_1.v_2}\nonumber\\
&&\left.+{\bf n.v_1} {\bf n.v_2} ({\bf v_1.v_2})^2-v_1^2 v_2^2 {\bf n.v_1} {\bf n.v_2}\right\}\,\nonumber,
\ee
where ${\bf n}\equiv{\bf r}/r$ and the Newton's $G_N$ constant is related to the $d-$dimensional gravitational constant $G$ by the following relation involving the arbitrary subtraction lenght scale $L$:
\be
\label{GtoGN}
G=G_N L^{d-3}\,,
\ee
which gives trivially $G\rightarrow G_N$ in this case, but gives also a nontrivial contribution in presence of divergences, see next section.

\subsection{${\cal O}(G^2)$ diagrams}
\begin{figure}
\includegraphics[width=1.\linewidth]{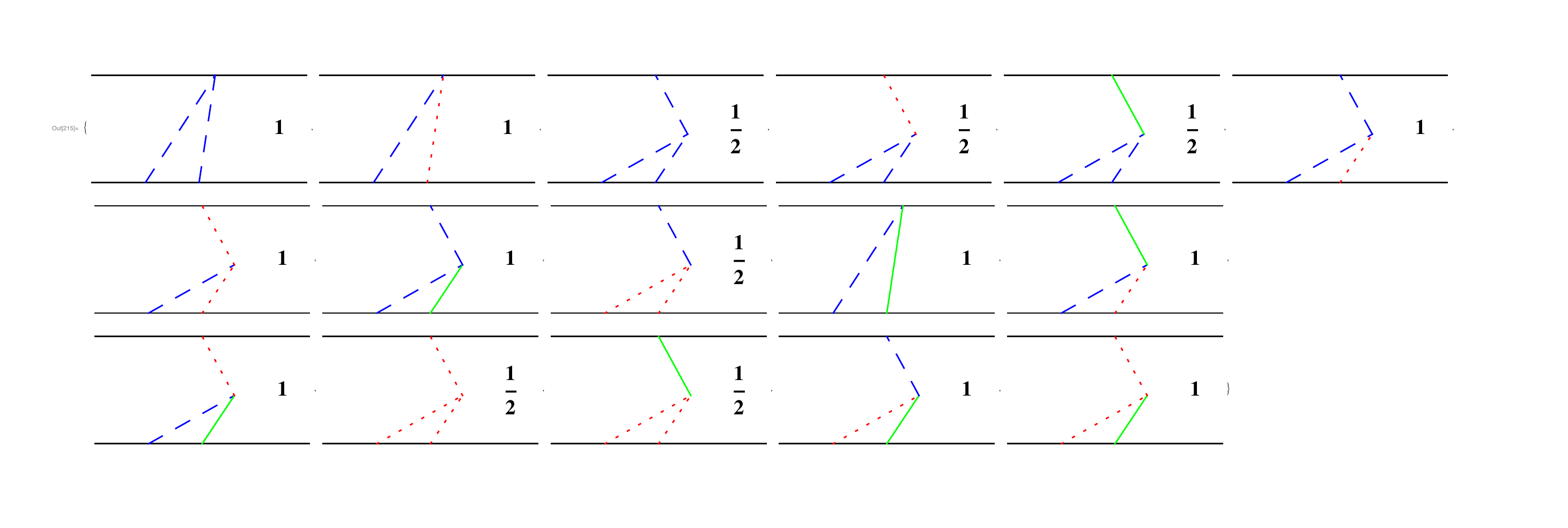}
\caption{The 16 ${\cal O}(G^2)$ diagrams.
Diagrams from 2 to 9 actually
start contributing at 2PN order, and must be consequently evaluated at their next-to leading order in $v^2$
(which involves an appropriate vertices expansion or one double time derivative insertion in the propagators)
to catch their 3PN contribution. Analogously, the first diagram enters already at 1PN order, so it must be expanded up to $v^4$ .
As in fig.~\ref{diaG1}, 
the $\phi$, $A$ and $\sigma$ 
propagators are represented respectively by blue dashed, red dotted and 
green solid lines and the number in the right part of each diagram is its multiplicity 
factor.}
\label{diaG2}
\end{figure}

The first, second and tenth diagram of figure \ref{diaG2} can be factorized in terms of lowest order
ones, all the others need the use of the well known formula
\be
\label{itkz}
\ds\int_{\bf p_1}\, \frac{1}{p_1^{2\alpha}({\mathbf p-{\bf p_1}})^{2\beta}}=
\frac 1{\pa{4\pi}^{d/2}}
\frac{\Gamma(d/2-\beta)\Gamma(d/2-\alpha)\Gamma(\alpha+\beta-d/2)}
{\Gamma(\alpha)\Gamma(\beta)\Gamma(d-\alpha-\beta)} p^{d-2\alpha-2\beta}\,,
\ee
and of its generalization to inclusion of $p_1^\mu$'s factors in the integrand
(up to four different indices are needed at 3PN), which can be worked out by means of appropriate contractions with $p^\mu$'s.
The result is:
\be
G^{2}_{1}&=&-\frac{G_N^2 m_1^2 m_2 }{16 r^2}\left[2 r^2 {\bf a_1.a_2}+36 v_2^2 {\bf r.a_1}+8 {\bf v_1.a_2} {\bf r.v_1}-8 {\bf r.a_2} ({\bf n.v_1})^2-8  ({\bf v_1.v_2})^2-24 v_1^2 {\bf v_1.v_2}\right.\nonumber\\
&&+144 v_1^2 v_2^2-32 v_1^4+49 v_2^4+48 v_1^2{\bf n.v_1} {\bf n.v_2}-24 ({\bf n.v_1})^2 {\bf v_1.v_2}+32 {\bf n.v_1} {\bf n.v_2} {\bf v_1.v_2}\nonumber\\
&&\left.-28 v_2^2 ({\bf n.v_1})^2-24 v_1^2 ({\bf n.v_1})^2-32 ({\bf n.v_1})^2 ({\bf n.v_2})^2+32 ({\bf n.v_1})^3 {\bf n.v_2}-8 ({\bf n.v_1})^4\right]\,,\nonumber\\
G^{2}_{2}&=&\frac{2 G_N^2 m_1^2 m_2}{r^2} \left[r^2 {\bf a_1.a_2}+2 {\bf v_1.a_2} {\bf r.v_1}-2({\bf v_1.v_2})^2-4 v_1^2 {\bf v_1.v_2}+3 v_2^2 {\bf v_1.v_2}-2 ({\bf n.v_1})^2 {\bf v_1.v_2}\right.\nonumber\\
&&\left.+4 {\bf n.v_1} {\bf n.v_2} {\bf v_1.v_2}\right]\,,\nonumber
\ee
\be
G^{2}_{3}&=&-\frac{G_N^2 m_1^2 m_2 }{3 (d-3)}[2 {\bf a_1.a_2}+a_1^2]+\frac{G_N^2 m_1^2 m_2}{36 r^2} \left[24 r^2 \log \left(\frac{r}{L_0}\right)\left(2{\bf a_1.a_2} + a_1^2 \right)-r^2 \left(32 {\bf a_1.a_2}+22 a_1^2\right)\right.\nonumber\\
&&+48 {\bf r.a_1} {\bf r.a_2}-180 v_1^2 {\bf r.a_1}-96 {\bf v_2.a_1} {\bf r.v_2}-156 v_2^2 {\bf r.a_1}-12 ({\bf r.a_1})^2+96 {\bf r.a_1} ({\bf n.v_2})^2\nonumber\\
&&+54 {\bf v_1.a_2} {\bf r.v_1}+288 v_1^2 {\bf r.a_2}-54 {\bf r.a_2} ({\bf n.v_1})^2-54 ({\bf v_1.v_2})^2+150 v_1^2 {\bf v_1.v_2}-369 v_1^2 v_2^2-78v_1^4\nonumber\\
&&+576v_1^2 ({\bf n.v_2})^2-300 v_1^2 {\bf n.v_1} {\bf n.v_2}+42 ({\bf n.v_1})^2 {\bf v_1.v_2}+216 {\bf n.v_1} {\bf n.v_2} {\bf v_1.v_2}+243 v_2^2 ({\bf n.v_1})^2\nonumber\\
&&\left.+222 v_1^2 ({\bf n.v_1})^2-216 ({\bf n.v_1})^2 ({\bf n.v_2})^2-56 ({\bf n.v_1})^3 {\bf n.v_2}+20 ({\bf n.v_1})^4\right]\,,\nonumber
\ee
\be
G^{2}_{4}&=&\frac{G_N^2 m_1^2 m_2 }{6 r^2}\left[r^2 {\bf a_1.a_2}+3 {\bf r.a_1} {\bf r.a_2}+12 {\bf r.a_1} {\bf v_1.v_2}-3 {\bf v_2.a_1} {\bf r.v_2}-3  v_2^2 {\bf r.a_1}+6 {\bf r.a_1} ({\bf n.v_2})^2\right.\nonumber\\
&&+11 {\bf v_1.a_2} {\bf r.v_1}+25 v_1^2 {\bf r.a_2}-2 {\bf r.a_2} ({\bf n.v_1})^2-11 ({\bf v_1.v_2})^2+6 v_1^2{\bf v_1.v_2}-25v_1^2 v_2^2\nonumber\\
&&-3 v_2^2 {\bf v_1.v_2}+50v_1^2 ({\bf n.v_2})^2-24v_1^2 {\bf n.v_1} {\bf n.v_2}-12 ({\bf n.v_1})^2 {\bf v_1.v_2}+26 {\bf n.v_1} {\bf n.v_2} {\bf v_1.v_2}\nonumber\\
&&\left.+2 v_2^2 ({\bf n.v_1})^2+3 v_2^2 {\bf n.v_1} {\bf n.v_2}-8 ({\bf n.v_1})^2 ({\bf n.v_2})^2-4 ({\bf n.v_1})^3 {\bf n.v_2}\right]\,,\nonumber\\
G^{2}_{5}&=&\frac{G_N^2 m_1^2 m_2 }{12 r^2}\left[10 {\bf v_2.a_1} {\bf r.v_2}+2 v_2^2 {\bf r.a_1}-4 {\bf r.a_1} ({\bf n.v_2})^2+6 ({\bf v_1.v_2})^2+18v_1^2 v_2^2+3 v_2^4\right.\nonumber\\
&&\left.-24 v_1^2 ({\bf n.v_2})^2-12 {\bf n.v_1} {\bf n.v_2} {\bf v_1.v_2}-3 v_2^2 ({\bf n.v_2})^2+12 ({\bf n.v_1})^2 ({\bf n.v_2})^2\right]\,,\nonumber\\
G^{2}_{6}&=&-\frac{G_N^2 m_1^2 m_2}{3 r^2} \left[r^2 \left({\bf a_1.a_2}+2 a_1^2\right)+27v_1^2 {\bf r.a_1}-9 v_2^2 {\bf r.a_1}+14 {\bf v_1.a_2} {\bf r.v_1}+4 v_1^2 {\bf r.a_2}\right.\nonumber\\
&&-14 {\bf r.a_2} ({\bf n.v_1})^2-14 ({\bf v_1.v_2})^2-39  v_1^2 {\bf v_1.v_2}-22 v_1^2 v_2^2+9  v_1^4+8  v_1^2 ({\bf n.v_2})^2+78 v_1^2 {\bf n.v_1} {\bf n.v_2}\nonumber\\
&&\left.+56 {\bf n.v_1} {\bf n.v_2} {\bf v_1.v_2}+50 v_2^2 ({\bf n.v_1})^2-24 v_1^2 ({\bf n.v_1})^2-56 ({\bf n.v_1})^2 ({\bf n.v_2})^2+8 ({\bf n.v_1})^4\right]\,,\nonumber\\
G^{2}_{7}&=&\frac{8 G_N^2 m_1^2 m_2 }{d-3}{\bf a_1.a_2}+\frac{4 G_N^2 m_1^2 m_2}{r^2} \left[r^2 {\bf a_1.a_2} \left(1-4\log \left(\frac{r}{L_0}\right)\right)-2 {\bf r.a_1} {\bf v_1.v_2}\right.\nonumber\\
&&+4 {\bf v_2.a_1} {\bf r.v_2}-4 {\bf v_1.a_2} {\bf r.v_1}-  v_1^2 {\bf r.a_2}+4 ({\bf v_1.v_2})^2+3  v_1^2 {\bf v_1.v_2}+ v_1^2 v_2^2+ v_2^2 {\bf v_1.v_2}\nonumber\\
&&\left.-2 v_1^2 ({\bf n.v_2})^2+2  v_1^2 {\bf n.v_1} {\bf n.v_2}+2 ({\bf n.v_1})^2 {\bf v_1.v_2}-8 {\bf n.v_1} {\bf n.v_2} {\bf v_1.v_2}\right]\,,\nonumber\\
G^{2}_{8}&=&\frac{G_N^2 m_1^2 m_2 }{9 r^2}\left[9  v_1^2 {\bf r.a_1}-21  v_1^2 {\bf v_1.v_2}-27  v_1^2 v_2^2-33 v_1^4+42 v_1^2 {\bf n.v_1} {\bf n.v_2}+96 ({\bf n.v_1})^2 {\bf v_1.v_2}\right.\nonumber\\
&&\left.+54 v_2^2 ({\bf n.v_1})^2+24 v_1^2 ({\bf n.v_1})^2-128 ({\bf n.v_1})^3 {\bf n.v_2}+56 ({\bf n.v_1})^4\right]\,,\nonumber\\
G^{2}_{9}&=&\frac{4  G_N^2 m_1^2 m_2}{d-3}a_1^2+\frac{2 G_N^2 m_1^2 m_2 }{r^2}\left[4 r^2 a_1^2 \left( 1-\log \left(\frac{r}{L_0}\right)\right) +8 v_1^2  {\bf r.a_1}-6  v_1^2 {\bf v_1.v_2}\right.\nonumber\\
&&\left.-3 v_1^2 v_2^2+  v_1^4+12  v_1^2 {\bf n.v_1} {\bf n.v_2}-9  v_1^2({\bf n.v_1})^2\right]\,,\nonumber\\
G^{2}_{10}&=&-\frac{6 G_N^2 m_1^2 m_2 }{r^2}\left[({\bf {\bf v_1.v_2}})^2-v_1^2 v_2^2\right]\,,\nonumber\\
G^{2}_{11}&=&-\frac{8 G_N^2 m_1^2 m_2}{r^2} {\bf v_2.a_1} {\bf r.v_2}\,,\nonumber\\
G^{2}_{12}&=&-\frac{8 G_N^2 m_1^2 m_2 }{r^2}\left[{\bf v_1.a_2} {\bf r.v_1}-({\bf v_1.v_2})^2+v_1^2 {\bf v_1.v_2}-2 ({\bf n.v_1})^2 {\bf v_1.v_2}+2 {\bf n.v_1} {\bf n.v_2} {\bf v_1.v_2}\right]\,,\nonumber\\
G^{2}_{13}&=&\frac{4 G_N^2 m_1^2 m_2 }{r^2}\left[4{\bf r.a_1} {\bf v_1.v_2}+6 {\bf v_1.a_2} {\bf r.v_1}-v_1^2 {\bf r.a_2}-6 ({\bf v_1.v_2})^2+2  v_1^2 {\bf v_1.v_2}+v_1^2 v_2^2\right.\nonumber\\
&&\left.-2 v_1^2 ({\bf n.v_2})^2+v_1^2 {\bf n.v_1} {\bf n.v_2}-4 ({\bf n.v_1})^2 {\bf v_1.v_2}+12 {\bf n.v_1} {\bf n.v_2} {\bf v_1.v_2}\right]\,,\nonumber\\
G^{2}_{14}&=&\frac{2 G_N^2 m_1^2 m_2}{r^2} \left[v_1^2 v_2^2-2 ({\bf v_1.v_2})^2+v_1^2 ({\bf n.v_2})^2\right]\,,\nonumber\\
G^{2}_{15}&=&-\frac{8 G_N^2 m_1^2 m_2 }{r^2}v_1^2 {\bf r.a_1}\,,\nonumber\\
G^{2}_{16}&=&\frac{8 G_N^2 m_1^2 m_2 }{r^2}\left[v_1^2-2 ({\bf n.v_1})^2\right] {\bf v_1.v_2}\,\nonumber.
\ee
Divergences appear at this order from the terms $G^2_3$, $G^2_7$ and $G^2_9$, 
under the shape of $(d-3)$ simple poles, giving rise also 
to logarithmic factors involving the subtraction scale 
$L_0 \equiv L e^{-\gamma/2}/\sqrt{4\pi}$ , where $\gamma=.577216\dots$ is the 
Euler-Mascheroni constant and $L$ is the length scale written in eq.~(\ref{GtoGN}).
Following \cite{Blanchet_living}, in  sec.~\ref{3pnlag} we will get rid of the poles by means of a shift of the worldline parameters;
as to the logarithms involving the arbitrary subtraction scale, they are known to remain in the Lagrangian
and also in the equation of motion for ${\bf r}$, which is however a gauge-dependent variable.
They however drop out in the expression of observables, like the energy $E$ of the system as a function of the
orbital frequency $\omega$, which are gauge-independent quantities as both $E$ and $\omega$
can be measured asymptotically far from the binary system.

\subsection{${\cal O}(G^3)$ diagrams}
\begin{figure}
\includegraphics[width=1.\linewidth]{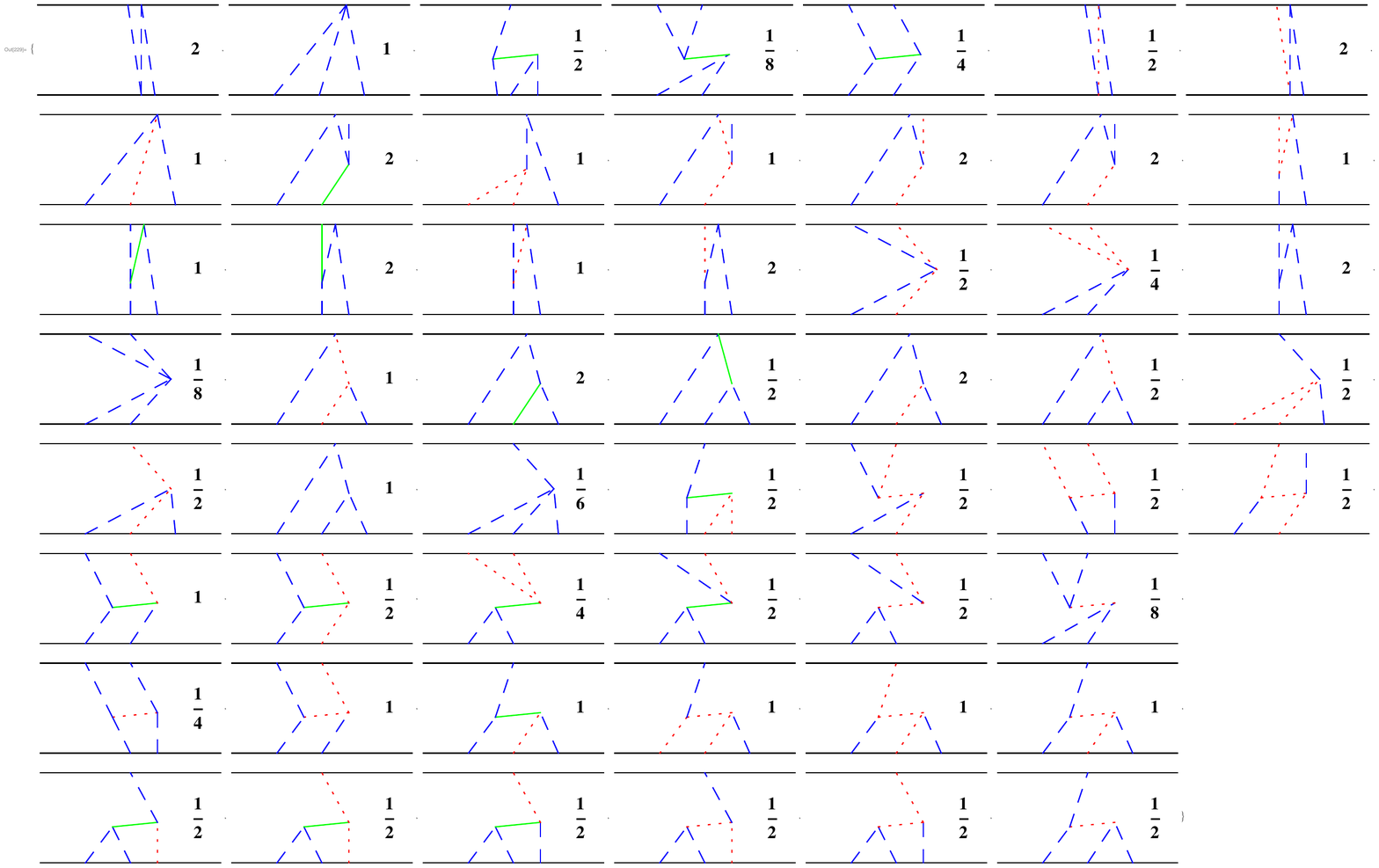}
\caption{The 53 ${\cal O}(G^3)$ diagrams. The first 5 enter at 2PN order, and must be consequently evaluated at their next-to
leading order in $v^2$, which involve an appropriate vertices expansion or one double time derivative insertion in the propagators.
As in fig.~\ref{diaG1} and 
\ref{diaG2}, the $\phi$, $A$ and $\sigma$ 
propagators are represented respectively by blue dashed, red dotted and 
green solid lines and the number in the right part of each diagram is its 
multiplicity factor.}
\label{diaG3}
\end{figure}

Most of the diagrams of figure \ref{diaG3} can be computed via recursive application of equation (\ref{itkz}), with
the exception of the 5th diagram and all the ones sharing the same H-shaped 
topology.
In this case one has to work on the integrand by parts to obtain the following
recursive relationship
\be
\label{rikparts}
\ba{rcl}
I(\alpha,\beta,\gamma,\delta,\epsilon) &\equiv &
\ds\int_{{\bf p_1},{\bf p_2}}\, \frac{1}{p_1^{2\alpha}
({\bf p}-{\bf p_1})^{2\beta}p_2^{2\gamma}({\bf p}-{\bf p_2})^{2\delta}
({\bf p_1}-{\bf p_2})^{2\epsilon}}\\
&=&\ds\left[\gamma\pa{I(\alpha-1,\beta,\gamma+1,\delta,\epsilon)-
I(\alpha,\beta,\gamma+1,\delta,\epsilon-1)}\right.+\\
&&\left.
\delta \pa{I(\alpha,\beta-1,\gamma,\delta+1,\epsilon)-
I(\alpha,\beta,\gamma,\delta+1,\epsilon-1)}\right]/
\pa{2\epsilon+\gamma+\delta-d}
\,,
\ea
\ee
and its generalization to the inclusion of $p_{1,2}^\mu$ factors in the integrand. 
The above formula (\ref{rikparts}) also holds exchanging in the right hand side 
($\{\alpha,\beta\}\leftrightarrow\{\gamma,\delta\}$) or 
($\{\alpha,\gamma\}\leftrightarrow\{\beta,\delta\}$) and 
it can be iterated to obtain an integral of the type eq.(\ref{itkz}).

As in the $G^2$ case, here also poles and logarithmic terms are obtained, 
which again are non physical, see sec.~\ref{3pnlag}.
The output of our code for these diagrams is:
\be
G^{3}_{1}&=&\frac{G_N^3 m_1^2 m_2^2}{4 r^3} \left[12 v_1^2-3 {\bf v_1.v_2}+7 {\bf n.v_1} {\bf n.v_2}-4 ({\bf n.v_1})^2\right]\,,\nonumber\\
G^{3}_{2}&=&-\frac{G_N^3 m_1^3 m_2 }{4 r^3}\left[{\bf v_1.v_2}+3 v_1^2+9 v_2^2-3 {\bf n.v_1} {\bf n.v_2}+2 ({\bf n.v_1})^2\right]\,,\nonumber\\
G^{3}_{3}&=&\frac{G_N^3 m_1^3 m_2} {30 (d-3) r^3}\left[10 {\bf v_1.v_2}-13 v_1^2-30{\bf n.v_1} {\bf n.v_2}+39 ({\bf n.v_1})^2\right]\nonumber\\
&&+\frac{G_N^3 m_1^3 m_2 }{900 r^3}\left[90 \log\left(\frac{r}{L_0}\right)\left(13 v_1^2-10 {\bf v_1.v_2}+30 {\bf n.v_1} {\bf n.v_2} -39 ({\bf n.v_1})^2\right)\right.\nonumber\\
&&+\left. 850 {\bf v_1.v_2}-2401 v_1^2-450 v_2^2-3450 {\bf n.v_1} {\bf n.v_2}+3243 ({\bf n.v_1})^2\right]\,,\nonumber\\
G^{3}_{4}&=&-\frac{\pi ^2 G_N^3 m_1^2 m_2^2}{128 r^3} \left[{\bf v_1.v_2}+2 v_1^2-3 {\bf n.v_1} {\bf n.v_2}-6 ({\bf n.v_1})^2\right]\,,\nonumber
\ee
\be
G^{3}_{5}&=&\frac{G_N^3 m_1^2 m_2^2}{72 r^3} \left[9\pi ^2 \left(v_1^2-{\bf v_1.v_2}+3 {\bf n.v_1} {\bf n.v_2}-3({\bf n.v_1})^2\right)\right.\nonumber\\
&&+\left. 50 {\bf v_1.v_2}-644 v_1^2-294 {\bf n.v_1} {\bf n.v_2}+348 ({\bf n.v_1})^2\right]\,,\nonumber\\
G^{3}_{6}&=& \frac{G^{3}_{7}}{2}=-\frac{G^{3}_{11}}{4}=-\frac{G^{3}_{12}}{4}=\frac{2 G_N^3 m_1^2 m_2^2 }{r^3}{\bf v_1.v_2}\,,\nonumber\\
G^{3}_{8}&=&-\frac{G^{3}_{23}}{4}=-\frac{3 G^{3}_{29}}{16}=\frac{3 G^{3}_{46}}{32}=\frac{2 G_N^3 m_1^3 m_2}{r^3} {\bf v_1.v_2}\,,\nonumber\\
G^{3}_{9}&=&\frac{G^{3}_{34}}{8}=-\frac{G_N^3 m_1^2 m_2^2 }{r^3}\left[v_1^2-({\bf n.v_1})^2\right]\,,\nonumber\\
G^{3}_{10}&=& \frac{3 G^{3}_{28}}{4}=-\frac{3 G^{3}_{45}}{8}=\frac{4 G_N^3 m_1^3 m_2 }{r^3}v_1^2\,,\nonumber\\
G^{3}_{13}&=&\frac{2 G_N^3 m_1^2 m_2^2 }{r^3}\left[{\bf v_1.v_2}-2 {\bf n.v_1} {\bf n.v_2}+({\bf n.v_1})^2\right]\,,\nonumber
\ee
\be
G^{3}_{14}&=&\frac{8 G_N^3 m_1^2 m_2^2 }{r^3}v_1^2\,,\nonumber\\
G^{3}_{15}&=&-3 G^3_{16}=-\frac{6 G_N^3 m_1^2 m_2^2 }{r^3}\left[v_1^2-2 ({\bf n.v_1})^2\right]\,,\nonumber\\
G^{3}_{17}&=&-\frac{2 G_N^3 m_1^2 m_2^2}{r^3} \left[{\bf v_1.v_2}+v_1^2-2 {\bf n.v_1} {\bf n.v_2}-2 ({\bf n.v_1})^2\right]\,,\nonumber\\
G^{3}_{18}&=&\frac{2 G_N^3 m_1^2 m_2^2}{r^3} \left[4 ({\bf n.v_1})^2-{\bf v_1.v_2}-v_1^2 \right]\,,\nonumber\\
G^{3}_{19}&=&G^3_{20}=G^3_{33}=G^3_{39}=0\,,\nonumber\\
G^{3}_{21}&=&-\frac{G_N^3 m_1^2 m_2^2 }{r^3}\left[v_1^2-4{\bf v_1.v_2}+14 {\bf n.v_1} {\bf n.v_2}-7 ({\bf n.v_1})^2\right]\,,\nonumber\\
G^{3}_{22}&=&\frac{\pi ^2 G_N^3 m_1^2 m_2^2}{8 r^3} \left[v_1^2-4 {\bf v_1.v_2}+12 {\bf n.v_1} {\bf n.v_2}-3 ({\bf n.v_1})^2\right]\,,\nonumber\\
G^{3}_{24}&=&\frac{2 G_N^3 m_1^3 m_2}{r^3} \left[v_1^2-2 ({\bf n.v_1})^2\right]\,,\nonumber\\
G^{3}_{25}&=&\frac{3 G_N^3 m_1^3 m_2 }{2 r^3}\left[v_2^2-({\bf n.v_2})^2\right]\,,\nonumber\\
G^{3}_{26}&=&-\frac{2 G_N^3 m_1^3 m_2 }{r^3}\left[{\bf v_1.v_2}+ v_1^2-3 {\bf n.v_1} {\bf n.v_2}-({\bf n.v_1})^2\right]\,,\nonumber\\
G^{3}_{27}&=&\frac{G_N^3 m_1^3 m_2}{r^3} \left[ {\bf v_1.v_2}-{\bf n.v_1} {\bf n.v_2}\right]\,,\nonumber\\
G^{3}_{30}&=&-\frac{G_N^3 m_1^3 m_2 }{2 r^3}\left[v_1^2-4 {\bf v_1.v_2}+12 {\bf n.v_1} {\bf n.v_2}-5 ({\bf n.v_1})^2\right]\,,\nonumber\\
G^{3}_{31}&=&-\frac{4 G_N^3 m_1^3 m_2}{15 (d-3) r^3} \left[v_1^2-5 {\bf v_1.v_2}+15 {\bf n.v_1} {\bf n.v_2}-3 ({\bf n.v_1})^2\right]\nonumber\\
&&-\frac{2 G_N^3 m_1^3 m_2}{75 r^3} \left[30\log\left(\frac{r}{L_0}\right) \left(5{\bf v_1.v_2}-v_1^2 -15 {\bf n.v_1} {\bf n.v_2} +3 ({\bf n.v_1})^2 \right)\right.\nonumber\\
&&\left.-25 {\bf v_1.v_2}- (v_1^2)+225 {\bf n.v_1} {\bf n.v_2}-57 ({\bf n.v_1})^2\right]\,,\nonumber\\
G^{3}_{32}&=&\frac{2 G_N^3 m_1^3 m_2}{r^3} \left[v_1^2-3 ({\bf n.v_1})^2\right]\left[\frac{2}{3 (d-3)}-2 \log\left(\frac{r}{L_0}\right)+1\right]\,,\nonumber\\
G^{3}_{35}&=&\frac{8 G_N^3 m_1^2 m_2^2}{r^3} \left[{\bf v_1.v_2}-{\bf n.v_1} {\bf n.v_2}\right]\,,\nonumber\\
G^{3}_{36}&=&-\frac{G_N^3 m_1^2 m_2^2 }{2 r^3}\left[3\pi ^2\left(v_1^2-{\bf v_1.v_2}+3{\bf n.v_1} {\bf n.v_2}-3 ({\bf n.v_1})^2\right)\right.\nonumber\\
&&\left.+ 60 {\bf v_1.v_2}-60 v_1^2-148 {\bf n.v_1} {\bf n.v_2}+148 ({\bf n.v_1})^2\right]\,,\nonumber\\
G^{3}_{37}&=&-\frac{G_N^3 m_1^2 m_2^2 }{2 r^3}\left[\pi ^2\left( {\bf v_1.v_2}-3 {\bf n.v_1} {\bf n.v_2}\right)-32{\bf v_1.v_2}+32 {\bf n.v_1} {\bf n.v_2}\right]\,,\nonumber\\
G^{3}_{38}&=&\frac{\pi ^2 G_N^3 m_1^2 m_2^2}{8 r^3} \left[ v_1^2-3 ({\bf n.v_1})^2\right]\,,\nonumber\\
G^{3}_{40}&=&-\frac{32 G^{3}_{41}}{3}=-\frac{\pi ^2 G_N^3 m_1^2 m_2^2 }{4 r^3}\left[{\bf v_1.v_2}-3 {\bf n.v_1} {\bf n.v_2}\right]\,,\nonumber\\
G^{3}_{42}&=&\frac{G_N^3 m_1^2 m_2^2 }{4 r^3}\left[{\bf v_1.v_2}+2 v_1^2+{\bf n.v_1} {\bf n.v_2}+2 ({\bf n.v_1})^2\right]\,,\nonumber\\
G^{3}_{43}&=&\frac{G_N^3 m_1^2 m_2^2}{2 r^3} \left[\pi^2 \left({\bf v_1.v_2}+v_1^2-3 {\bf n.v_1} {\bf n.v_2}-3 ({\bf n.v_1})^2\right)+4 {\bf v_1.v_2}-4 v_1^2+4 {\bf n.v_1} {\bf n.v_2}-4 ({\bf n.v_1})^2\right]\,,\nonumber
\ee
\be
G^{3}_{44}&=&\frac{4 G_N^3 m_1^3 m_2}{(d-3) r^3} \left[v_1^2-{\bf v_1.v_2}+3 {\bf n.v_1} {\bf n.v_2}-3 ({\bf n.v_1})^2\right]\nonumber\\
&&+\frac{2 G_N^3 m_1^3 m_2}{r^3} \left[6 \log\left(\frac{r}{L_0}\right) \left({\bf v_1.v_2}-v_1^2-3 {\bf n.v_1} {\bf n.v_2}+3 ({\bf n.v_1})^2\right)\right.\nonumber\\
&&\left.+5 v_1^2-5 {\bf v_1.v_2}+21 {\bf n.v_1} {\bf n.v_2}-21 ({\bf n.v_1})^2\right]\,,\nonumber\\
G^{3}_{47}&=&-\frac{4 G_N^3 m_1^3 m_2 }{3 (d-3) r^3}\left[{\bf v_1.v_2}+v_1^2-3 {\bf n.v_1} {\bf n.v_2}-3 ({\bf n.v_1})^2\right]\nonumber\\
&&+\frac{2 G_N^3 m_1^3 m_2 }{3 r^3}\left[6 \log\left(\frac{r}{L_0}\right) \left(v_1^2+{\bf v_1.v_2}-3{\bf n.v_1} {\bf n.v_2} -3 ({\bf n.v_1})^2\right)\right.\nonumber\\
&&\left.-{\bf v_1.v_2}-v_1^2+9 {\bf n.v_1} {\bf n.v_2}+9 ({\bf n.v_1})^2\right]\,,\nonumber\\
G^{3}_{48}&=&\frac{2 G_N^3 m_1^3 m_2 }{5 (d-3) r^3}\left[ v_1^2-3 ({\bf n.v_1})^2\right]+\frac{2 G_N^3 m_1^3 m_2}{25 r^3} \left[17 v_1^2-56 ({\bf n.v_1})^2- 15 \log\left(\frac{r}{L_0}\right)\left(v_1^2 -3 ({\bf n.v_1})^2\right)\right]\,,\nonumber\\
G^{3}_{49}&=&-\frac{4 G_N^3 m_1^3 m_2 }{15 (d-3) r^3}\left[{\bf v_1.v_2}-3 {\bf n.v_1} {\bf n.v_2}\right]-\frac{4 G_N^3 m_1^3 m_2 }{75 r^3}\left[12 {\bf v_1.v_2}-66 {\bf n.v_1} {\bf n.v_2}\right.\nonumber\\
&&\left.-15\log\left(\frac{r}{L_0}\right)\left( {\bf v_1.v_2}-3 {\bf n.v_1} {\bf n.v_2}\right)\right]\,,\nonumber\\
G^{3}_{50}&=&-\frac{2 G_N^3 m_1^3 m_2} {5 (d-3) r^3}\left[ {\bf v_1.v_2}-3 {\bf n.v_1} {\bf n.v_2}\right]-\frac{2 G_N^3 m_1^3 m_2}{25 r^3} \left[17 {\bf v_1.v_2}-56 {\bf n.v_1} {\bf n.v_2}\right.\nonumber\\
&&\left.-15\log\left(\frac{r}{L_0}\right) \left({\bf v_1.v_2}-3{\bf n.v_1} {\bf n.v_2} \right)\right]\,,\nonumber\\
G^{3}_{51}&=&-\frac{2 G_N^3 m_1^3 m_2} {15 (d-3) r^3}\left[ v_1^2-3 ({\bf n.v_1})^2\right]+\frac{2 G_N^3 m_1^3 m_2}{75 r^3} \left[23 v_1^2+36 ({\bf n.v_1})^2\right.\nonumber\\
&&\left.+15\log\left(\frac{r}{L_0}\right) \left(v_1^2-3({\bf n.v_1})^2 \right)\right]\,,\nonumber\\
G^{3}_{52}&=&\frac{2 G_N^3 m_1^3 m_2 }{3 (d-3) r^3}\left[{\bf v_1.v_2}-3 {\bf n.v_1} {\bf n.v_2}\right]-\frac{2 G_N^3 m_1^3 m_2 }{3 r^3}\left[ {\bf v_1.v_2}+6 {\bf n.v_1} {\bf n.v_2}\right.\nonumber\\
&&\left.+3 \log\left(\frac{r}{L_0}\right)\left( {\bf v_1.v_2}-3 {\bf n.v_1} {\bf n.v_2} \right)\right]\,,\nonumber\\
G^{3}_{53}&=&\frac{G_N^3 m_1^3 m_2 }{10 (d-3) r^3}\left[ v_1^2-3 ({\bf n.v_1})^2\right]+\frac{G_N^3 m_1^3 m_2}{300 r^3} \left[17 v_1^2-100 {\bf v_1.v_2}-81 ({\bf n.v_1})^2+300 {\bf n.v_1} {\bf n.v_2}\right.\nonumber\\
&&\left.-90 \log\left(\frac{r}{L_0}\right)\left(v_1^2 -3 ({\bf n.v_1})^2 \right)\right]\,\nonumber.
\ee

\subsection{${\cal O}(G^4)$ diagrams}
The $G^4$ topology set shown in figure \ref{diaG4} does not present any new difficulty, and the 
integrations can be done straightforwardly:
\begin{center}
  \begin{figure}
    \includegraphics[width=1.\linewidth]{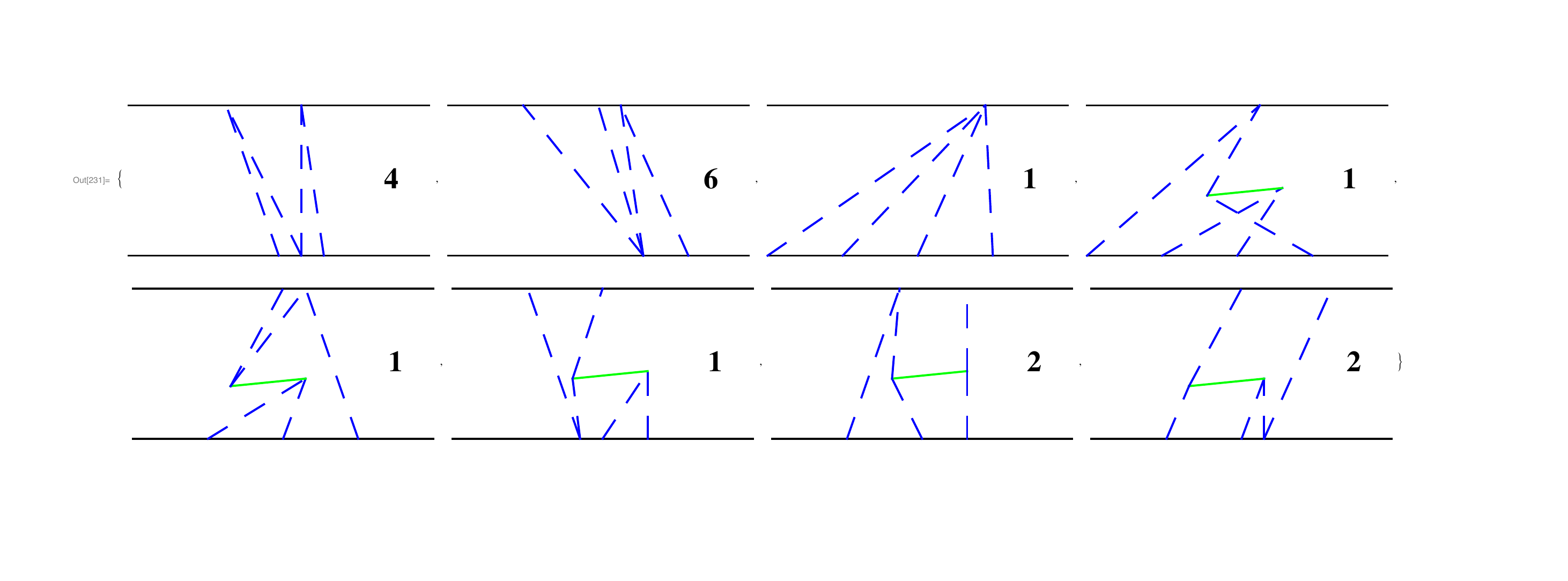}
    \caption{The 8 ${\cal O}(G^4)$ diagrams. As in fig.~\ref{diaG1}, \ref{diaG2}
      and \ref{diaG3}, the $\phi$, $A$ and $\sigma$ 
      propagators are represented respectively by blue dashed, red dotted and 
      green solid lines. 
      The number in the right part of each diagram is its multiplicity factor.}
    \label{diaG4}
  \end{figure}
\end{center}
\be
G^{4}_1&=&G^4_2=\frac{3G^4_6}{2}=\frac{G^4_7}{8}=\frac{3G^4_8}{4}=\frac{G_N^4 m_1^3 m_2^2}{2 r^4}\,,\nonumber\\
G^{4}_3&=&\frac{G^4_4}{8}=\frac{G_N^4 m_1^4 m_2}{24 r^4}\,,\nonumber\\
G^{4}_5&=&0\,.\nonumber
\ee

\subsection{Reproducing the 3PN Lagragian}
\label{3pnlag}
Adding up all the terms shown so far a Lagrangian containing
poles and logarithmic terms is obtained.
The poles can be removed by performing the shift of the worldline parameters as
given by eq.(1.13) of \cite{Blanchet:2003gy}, acting over the source positions 
${\bf x_{1,2}}$ according to ${\bf x_1}\to {\bf x_1'}={\bf x_1}+{\mathbf{\xi_1}}$
with
\be
{\mathbf{\xi_1}}\equiv \frac{11}3 G^2m_1^2\paq{\frac 1{d-3}-
2\log\pa{\frac r{L_0}}-\frac{327}{1540}}{\bf a_1}\,,
\ee
and analogously for ${\mathbf \xi_2}$
\footnote{Since the transformation is of ${\cal O}$(3PN), it should be implemented on the Newtonian part of the Lagrangian only,
with the caution that also the ${\cal O}(d-3)$ part of the Newtonian potential should be taken into account, as it gives a nonvanishing finite part
under the action of the divergent part of the shift of the worldline 
parameters.}.

Finally, in order to get the exact form of the 3PN Lagrangian as it is written in eq.~(174) of \cite{Blanchet_living} (and
originally derived in \cite{deAndrade:2000gf}),
one has to act on factors like ${\bf a_1}\cdot {\bf a_2}$, ${\bf \dot{a}_1}\cdot {\bf \dot{a}_2}$ and similar
by means of the standard \emph{double-zero} technique,
see e.g.~\cite{multizero},
\renewcommand{\arraystretch}{1.4}
\be
\label{doubzer}
\ba{rcl}
{\bf a_1}\cdot {\bf a_2} &=& \ds
\pa{{\bf a_1}+\frac{G m_2}{r^3}\R-\frac{Gm_2}{r^3}\R}\cdot
\pa{{\bf a_2}-\frac{G m_1}{r^3}\R+\frac{Gm_1}{r^3}\R}\\
&\simeq & \ds {\bf a_1}\cdot \frac{Gm_1}{r^3}\R-
{\bf a_2}\cdot\frac{Gm_2}{r^3}\R+\frac{G^2m_1m_2}{r^4}\,,
\ea
\ee
\renewcommand{\arraystretch}{1}
which makes use of the lowest order equations of motion to get rid of 
the 3PN Lagrangian terms quadratic in the accelerations.
This transformation should be iterated (as in the case of the ${\bf \dot{a}_1}\cdot {\bf \dot{a}_2}$ term) and combined with derivations
by parts until only terms linear in the accelerations remain.

Moreover, as terms like ${\bf a_1}\cdot {\bf a_2}$ appear also in the 2PN effective Lagrangian, the same technique must be applied 
at 2PN, but employing the equation of motion up to 1PN accuracy rather than just the Newtonian part written above;
this gives a further contribution to the 3PN Lagrangian, and all these terms nicely fit with all the previous ones
to give eq. (174) of \cite{Blanchet_living}.

\section{Conclusions}
\label{conclusion}
We have computed the conservative dynamics of a gravitationally bound binary system within the
framework of the PN approximation to general relativity to third post-Newtonian
order.
By a systematic use of the effective field theory methods for general 
relativity proposed by Goldberger
and Rothstein it has been possible to automatize the computation and compute
the effective Lagrangian by summing the contributions of eighty Feynman diagrams.
We have computed the effective Lagrangian reproducing known results
\cite{energy_at_3PN} and paving the way for the yet uncomputed 4PN dynamics.
Beside the clear theoretical interest of this calculation, the 4PN Hamiltonian 
has a direct phenomenological impact as it enters the determination of the 
templates waveforms used in analyzing the data from the large interferometers 
LIGO and Virgo like the Effective One Body ones \cite{EOB}.

\section*{Acknowledgments}
It is a pleasure to thank Michele Maggiore for stimulating discussions and Luc 
Blanchet and Walter Goldberger for useful correspondance.
RS wishes to thank the theoretical physics department of the University of 
Geneva for kind hospitality and support during the preparation of part this 
work.
SF wishes to thank the Science Department of the University of Urbino for 
hospitality during the preparation of part of this work.
The work of SF is supported by the FNS, the work of RS is supported
by the EGO Consortium through the VESF fellowship EGO-DIR-41-2010.

\end{document}